\documentclass[twocolumn,floats,showpacs,amsmath,amssymb,prb]{revtex4}
\usepackage{graphicx}
\begin{document}
                                                                                   
\title{Molecular hydrogen in silicon: A path-integral simulation}
\author{Carlos P. Herrero}
\author{Rafael Ram\'{\i}rez}
\affiliation{Instituto de Ciencia de Materiales,
         Consejo Superior de Investigaciones Cient\'{\i}ficas (CSIC),
         Campus de Cantoblanco, 28049 Madrid, Spain }
\date{\today}

\begin{abstract}
   Molecular hydrogen in silicon has been studied 
by path-integral molecular dynamics simulations in the canonical ensemble.
 Finite-temperature properties of these point defects were analyzed in 
the range from 300 to 900 K.
Interatomic interactions were modeled by a tight-binding potential
fitted to density-functional calculations. 
The most stable position for these impurities 
is found at the interstitial T site, with the hydrogen molecule rotating
freely in the Si cage. 
Vibrational frequencies have been obtained from a linear-response approach,
based on correlations of atom displacements at finite temperatures.
The results show a large anharmonic effect in the stretching vibration,
$\omega_s$, which is softened with respect to a harmonic approximation 
by about 300 cm$^{-1}$. 
The coupling between rotation and vibration causes an important decrease
in $\omega_s$ for rising temperature.
\end{abstract}


\maketitle

\section{Introduction}
Hydrogen can be incorporated into semiconductors both intentionally 
and unintentionally during manufacturing processes carried out for
technological applications.
It appears in these solids in a number of different 
configurations: as an isolated interstitial, bound to impurities, 
bound to native defects, in molecular form, etc.\cite{pe92,es95,st91} 
In the early 1980s, isolated hydrogen molecules were predicted to 
be stable in crystalline semiconductors and to play an important role 
in the diffusion of hydrogen in these materials.\cite{ma83,co83}
However, they were not unambiguously detected by spectroscopic
methods until more than ten years later.\cite{ve96,pr97,le98}

Vibrational transitions have been reported for interstitial H$_2$
in Si,\cite{pr97,le98} Ge,\cite{hi05} and GaAs.\cite{ve96} 
In these semiconductors, theory predicts that the
H$_2$ molecule is stable at an interstitial tetrahedral (T) site 
and behaves as a nearly free rotator.\cite{ok97,wa98}  
This gives rise at low temperatures to two stretching local vibrational
modes originating from para and ortho nuclear states, which are split
due to ro-vibrational coupling.\cite{ch02b}
                                                                                    
Here we will concentrate on isolated hydrogen molecules in the bulk
of crystalline silicon.
The interest of this problem is twofold. On one side, it is important
as a point defect in semiconductor physics, for its relevance in 
the hydrogen diffusion and stability in these materials.
On the other side, from a fundamental point of view, H$_2$ in silicon 
is an example of a light molecule sitting and moving in a confined geometry,
and one can study its behavior when localized in a spatial region with
extension of a few \AA. 

Earlier theoretical studies of molecular hydrogen in semiconductors
have concentrated on determining the lowest-energy site and stretching
frequency of the molecule, including in some cases anharmonic effects
derived from the calculated potential-energy 
surface,\cite{ok97,ok98,ho98,wa98,pr02}
as well as the quantum rotation of H$_2$ molecules.\cite{fo02,ho03}
Density-functional electronic-structure calculations in condensed matter 
are very reliable, but they treat atomic nuclei as classical particles,
and typical quantum effects like zero-point vibrations are 
not directly accessible.
These effects can be included by employing harmonic or
quasiharmonic approximations, but are difficult to take into account
when large anharmonicities are present, as can happen for
light impurities like hydrogen.

To consider the quantum character of the nuclei,
the path-integral molecular dynamics (or Monte Carlo) approach has
proved to be very useful.
A remarkable advantage of this method is that all nuclear degrees of
freedom can be quantized in an efficient manner, thus including
both quantum and thermal fluctuations in many-body systems
at finite temperatures.  In this way, Monte Carlo or molecular dynamics 
sampling applied to evaluate finite-temperature path integrals allows one to 
carry out quantitative and nonperturbative studies of highly-anharmonic 
effects in solids.\cite{gi88,ce95}

In this paper, the path-integral molecular dynamics (PIMD) method is used 
to study interstitial hydrogen molecules in silicon. 
Special attention has been paid to the vibrational properties of these
impurities, by considering anharmonic effects on their quantum dynamics and
the ro-vibrational coupling at different temperatures.
The results of the present calculations show that anharmonic effects lead
to a significant decrease of the vibrational frequencies of the 
impurities, as compared to a harmonic approximation.  
We have analyzed the isotopic effect on structural and vibrational
properties of these molecules, by considering also molecular deuterium (D$_2$). 
Path-integral methods analogous to that employed in this work 
have been applied earlier to study hydrogen in metals\cite{gi88} 
and semiconductors,\cite{ra94,he95,mi98,he06,he07} as well as on 
surfaces.\cite{ma95,he09}
In connection with the behavior of molecular hydrogen in confined regions, 
H$_2$ has been studied inside carbon nanotubes by diffusion Monte 
Carlo.\cite{go01}
Also, path-integral simulation methods have been extensively applied 
to study condensed phases of hydrogen in molecular 
form.\cite{ka94,su97,ch99,ki00}

 The paper is organized as follows. In Sec.\,II, we describe the
computational method and the models employed in our calculations. 
Our results are presented in Sec.\,III, dealing with the kinetic energy 
of the molecules, spatial delocalization, interatomic distance, and 
vibrational frequencies.
Sec.\,IV includes a discussion of the results and a summary.

\section{Computational Method}

\subsection{Path-integral molecular dynamics}

In the path-integral formulation of statistical mechanics employed here, 
the partition function is evaluated by a discretization of the density matrix
along cyclic paths, consisting of a finite number $P$ (Trotter number)
of ``imaginary-time'' steps.\cite{fe72,kl90} In the implementation in 
numerical simulations, this discretization gives rise to the appearance of 
$P$ ``beads'' for each quantum particle. These beads can be formally
treated as classical particles, so that the partition function of the
original quantum system is isomorph to that of a classical one. 
This isomorphism is obtained by replacing each quantum particle by a 
ring polymer consisting of $P$ classical particles, connected by harmonic 
springs.\cite{gi88,ce95}
In many-body problems, the configuration space is usually sampled by
Monte Carlo or molecular dynamics techniques. Here, we have employed the
PIMD method, which has been found to need less computer time for the
present problem.
We have used effective algorithms for performing PIMD simulations 
in the canonical $NVT$ ensemble, as those described in detail by 
Martyna {\em et al.}\cite{ma96} and Tuckerman.\cite{tu02}

Our calculations have been performed within the adiabatic
(Born-Oppenheimer) approximation, which allows one to define a potential
energy surface for the nuclear motion.
An important issue in this kind of simulations is the proper description 
of interatomic interactions, which should be as realistic as possible.
Since using true density functional (DF) or Hartree-Fock-type calculations
requires computer resources that would restrict enormously the size of
our simulation cell, we obtain the Born-Oppenheimer surface from
a tight-binding (TB) effective Hamiltonian, derived from DF 
calculations.\cite{po95}
The TB energy consists of two parts, the first one is the sum of
energies of occupied one-electron states, and the second one is 
given by a pairwise repulsive interatomic potential.\cite{po95}
For the present study the H-H pair potential was tuned to reproduce
the main features of known effective interatomic potentials, such
as the Morse potential.\cite{ra01} 
The capability of TB methods to simulate different properties of
solids and molecules has been reviewed by Goringe {\em et al.}\cite{go97}
The convergence of the total energy with the sampling in reciprocal space
was checked by using several sets of special $k$-points.\cite{ra88}
We found that a set of 4 $k$-points provides already good convergence
(relative error less than 0.001 \% in the total energy).
The use of only the $\Gamma$ point introduces a small systematic
error in the total energy that affects slightly the value of energy 
differences between different spacial configurations of H$_2$ in silicon, 
with typical errors of about 0.01 eV.
These results justify that the simulations presented in this work were
performed by using only the $\Gamma$ point for the reciprocal space sampling.

  Simulations were carried out on a $2\times2\times2$ supercell of the 
silicon face-centered cubic cell with periodic boundary conditions,
containing 64 Si atoms and a hydrogen (or deuterium) molecule. 
For comparison, we also 
carried out simulations of pure silicon, using the same supercell size.
Sampling of the configuration space has been carried out 
at temperatures between 300 and 900 K. 
The electronic structure calculations were performed without considering a
temperature-dependent Fermi filling of the electronic states, which is
reasonable for this temperature range.
 For a given temperature, a typical simulation run consisted of $10^4$ PIMD 
steps for system equilibration, followed by $5 \times 10^5$ steps for the 
calculation of ensemble average properties.
To keep a nearly constant precision in the path integral results
at different temperatures, we have employed a Trotter number that
scales as the inverse temperature. In particular, we have taken 
$P T$ = 18000 K, which means $P$ = 60 for  $T$ = 300 K.
Quantum exchange effects between protons or deuterons were not considered,
as they are negligible at the temperatures considered here, so that both
atomic nuclei in a molecule were treated as if they were distinguishable
particles.

The simulations were carried out by employing a staging transformation
for the bead coordinates.
The canonical ensemble was generated by coupling chains of four 
Nos\'e-Hoover thermostats (with mass $Q = \beta \hbar^2 / 5 P$)
to each degree of freedom.\cite{tu98}
To integrate the equations of motion, we used
a reversible reference-system propagator algorithm (RESPA), which allows
one to define different time steps for the integration of fast and slow
degrees of freedom.\cite{ma96} 
The time step $\Delta t$ associated to the calculation of DF-TB forces
was taken in the range
between 0.1 and 0.4 fs, which was found to be appropriate for the
interactions, atomic masses, and temperatures under consideration.
For the evolution of the fast dynamical variables, including the
thermostats and harmonic bead interactions, we used a smaller
time step $\delta t = \Delta t/4$.
We note that for H$_2$ in silicon at 300 K, a simulation run consisting of 
$5 \times 10^5$ PIMD steps needs the calculation 
of forces and energy with the TB code for 
$3 \times 10^7$ configurations, which has required the use of parallel 
computers.

\subsection{Calculation of anharmonic vibrational frequencies}

  Vibrational frequencies of impurities in solids are important
characteristics, which depend on the site that they actually occupy 
and on its interactions with the nearby hosts atoms.
In this context, the question arises whether the oscillator 
frequencies associated to an impurity can be extracted by assuming the 
host atoms fixed in the relaxed geometry corresponding to the 
minimum-energy configuration. 
This is a method usually employed to calculate vibrational frequencies 
of impurities in crystals.
On the other side, when the host atoms are allowed to relax by following 
the impurity motion, the potential energy surface is flatter 
than when the host atoms are fixed. 
To obtain an approach for the actual vibrational frequencies of the
impurities, one can calculate the eigenvalues of the dynamical matrix
of the whole simulation cell, and obtain the frequencies in the harmonic 
approximation (HA). 
However, for light impurities the anharmonicity can be appreciable, and 
the harmonic frequencies are only a first (maybe crude) approximation.

To calculate anharmonic frequencies we will use here a method based 
on the linear response (LR) of the system to vanishingly small forces applied 
on the atomic nuclei.
To this end, we consider a LR function, the static isothermal 
susceptibility $\chi^T$, that is readily derived from PIMD simulations of the
equilibrium solid, without dealing explicitly with any external forces in 
the simulation. This approach represents a significant improvement as
compared to a standard harmonic approximation.\cite{ra01}
The tensor ${\chi}^T$ allows one to derive a LR approximation
to the low-lying excitation energies of the vibrational system,
that is applicable even to highly anharmonic situations.
For a system with $3 N$ vibrational degrees of freedom,
the LR approximation for the frequencies reads
\begin{equation}
\omega_n = \frac{1}{\sqrt{\delta_n}}   \;,
\end{equation}
where $\delta_n$ ($n = 1, \dots, 3N$) are eigenvalues of
$\chi^{T}$, and the LR approximation to the low-lying excitation energy of 
vibrational mode $n$ is given by $\hbar\omega_n$. 
Details on the method and illustrations of its ability for predicting
vibrational frequencies of solids and molecules are given
elsewhere.\cite{ra01,ra02,lo03,ra05}

\section{Results}

\subsection{Minimum-energy configuration}

We first present results for classical calculations at
zero temperature, where the atoms are treated as point-like
particles without spatial delocalization.
The employed interatomic potential gives
reliable results for molecular hydrogen in vacuo.
The lowest-energy molecular configuration corresponds to a
distance $R_0$ between hydrogen atoms of 0.741 \AA. 
At this distance we obtain for H$_2$ in a harmonic approximation 
a stretching frequency of 4397 cm$^{-1}$. 

For H$_2$ as an impurity in silicon, we find a
lowest-energy position for the center-of-gravity of the molecule
located at an interstitial T site.
The minimum energy is found for the H--H axis along a
$\langle 100 \rangle$ crystal direction, with a distance between 
H atoms of 0.752 \AA. Moreover, changes in the energy for molecule
rotation keeping its center-of-gravity at a T site are very small,
in agreement with earlier calculations based on DF 
theory.\cite{ok97,wa98,ho98}
In silicon an increase in the H--H distance of about 0.01 \AA\ was 
found with respect to the molecule in vacuo, as expected for
an attractive interaction between each H and the nearby Si atoms. 

Assuming the H$_2$ molecule at a T site, and oriented along the
$\langle 100 \rangle$ direction, we find in the harmonic approximation 
a stretching frequency of 4071 cm$^{-1}$,  close to the harmonic value
of 4015 cm$^{-1}$ derived from the (anharmonic) vibrational frequency 
observed in Raman spectra\cite{la02} This represents a decrease 
of more than 300 cm$^{-1}$ vs the harmonic frequency for the molecule in
the gas phase, in line with a weakening of the H--H bond due to
interaction with the silicon lattice, as discussed earlier.\cite{ok97,wa98}
In the HA we also find frequencies $\omega_{\|}$ = 954 cm$^{-1}$
and $\omega_{\perp}$ = 1385 cm$^{-1}$ (twofold degenerate) 
for motion of the molecule along and perpendicular to the H--H
axis in the silicon cage. These two vibrational frequencies are not
expected to be observable because they will be mixed by the free
rotation of the molecule (see below).

\subsection{Kinetic energy}

We now turn to our PIMD simulations at finite temperatures.
To obtain insight into the motion of H$_2$ around the tetrahedral
site, we will consider various models in which the number of
degrees of freedom will be successively increased. 
In particular, we will consider: (1) motion of the H$_2$ molecule 
in one dimension (along the H--H bond) in a fixed and unrelaxed 
silicon lattice;
(2) free motion (in 3d) of the hydrogen molecule with fixed
host atoms, and (3) free motion of H$_2$ with mobile Si atoms.
In the latter case, all 66 atomic nuclei in the simulation cell are 
treated as quantum particles.

In our finite-temperature simulations for cases (2) and (3), where
the molecule can rotate around the T site, we observe a free
molecular rotation, without any preferential orientation. This is in 
agreement with earlier conclusions derived from theoretical\cite{ok97,wa98}
and experimental\cite{ch02a,ch02b} works, and with the fact that the
potential-energy surface for the rotation does not display deep
minima. 

\begin{figure}
\vspace{-2.0cm}
\hspace{-0.5cm}
\includegraphics[width= 9cm]{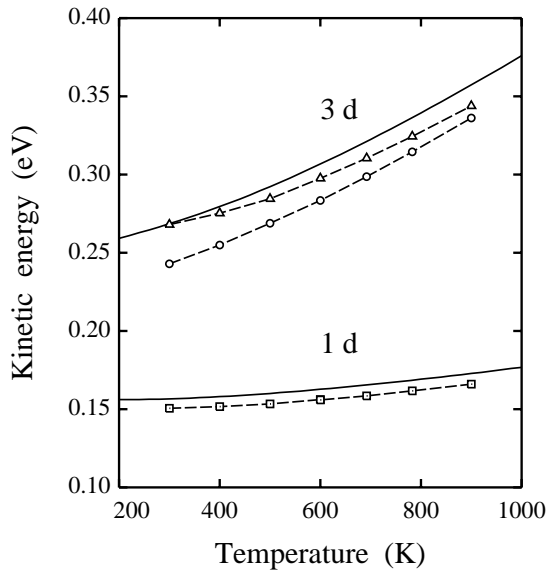}
\vspace{-2.5cm}
\caption{
Temperature dependence of the kinetic energy of molecular hydrogen
in silicon for various approximations.
Squares: motion of H$_2$ in one dimension with fixed host atoms;
circles: free motion of H$_2$ in a fixed silicon lattice;
triangles: free motion of H$_2$ with unrestricted motion of the
Si atoms.
Solid lines correspond to harmonic approximations for H$_2$ motion
in one and three dimensions.
Dashed lines are guides to the eye.
}
\label{f1}
\end{figure}

In Fig.~1 we show the kinetic energy of the hydrogen molecule in the
three considered approaches. Symbols indicate results derived from our PIMD
simulations using the so-called virial estimator\cite{he82,tu98} and solid 
lines represent the kinetic energy expected in a harmonic approximation. 
For 1d motion of H$_2$ (approach 1, squares) we find 
a slight increase in $E_k$ as temperature is raised.  In this approach,
results of the simulations are somewhat lower than those derived for
the HA, as expected for a softening of the vibrations due to the
anharmonicity of the interatomic potential.
In fact, the linear-response method introduced in Sect.~II.B  gives
in this case for the stretching frequency $\omega_s$ = 3770 cm$^{-1}$
at 300 K, 
which means a decrease of about 300 cm$^{-1}$ with respect to the 
harmonic model for H$_2$ in silicon ($\omega_s$ = 4071 cm$^{-1}$).

Circles in Fig.~1 correspond to our approach 2 with H$_2$ moving
in a fixed silicon lattice. Now we are dealing with six degrees of
freedom, two of which correspond to molecular rotation. This 
approximation gives again values of the kinetic energy smaller than 
those predicted by the HA (solid line).
This harmonic approximation includes a classical description of the
two rotational degrees of freedom of the H$_2$ molecule.
To analyze the kinetic energy of the defect complex in model 3
(all atoms are free to move), we calculate $E_k$ for the simulation
cell with and without the H$_2$ molecule:
$E_k$(defect) = $E_k$(64 Si + H$_2$) -- $E_k$(64 Si).
The results (triangles) lie appreciably above those obtained for model 2, 
indicating a nonnegligible coupling in the motion of interstitial molecule 
and host atoms.

\begin{figure}
\vspace{-2.0cm}
\hspace{-0.5cm}
\includegraphics[width= 9cm]{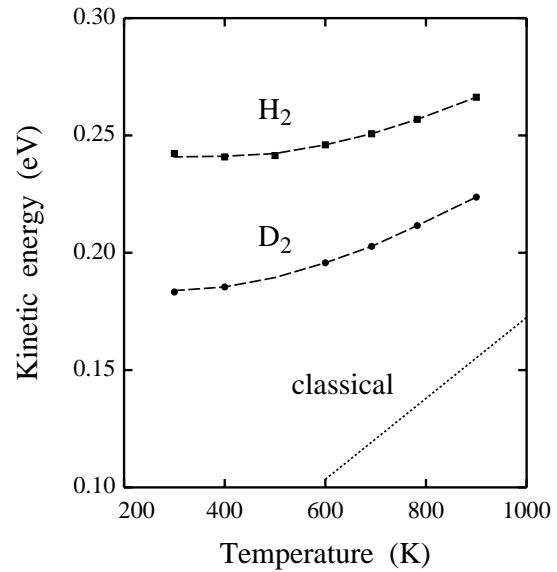}
\vspace{-2.5cm}
\caption{
Temperature dependence of the
vibrational part of the kinetic energy of H$_2$ and D$_2$, as
derived from approach 3 with free motion of all atoms in the
simulation cell.  Symbols indicate results derived from
PIMD simulations: squares for H$_2$ and circles for D$_2$.
Error bars are on the order of the symbol size.
Dashed lines are guides to the eye.
The dotted line corresponds to the classical limit with four
degrees of freedom.
}
\label{f2}
\end{figure}

At the temperatures considered here, rotation of H$_2$ can be
considered with a high precision to be classical. This means that 
its contribution to the kinetic energy of the molecule will be $k_B T$
(two degrees of freedom). 
Then, we can subtract this classical energy from $E_k$ derived
from the PIMD simulations to obtain a vibrational contribution to
the kinetic energy $E_k^v$. This part of the kinetic energy is shown in
Fig.~2 for H$_2$ (squares) and D$_2$ (circles). At low temperature
it converges to values close to 0.24 and 0.18 eV, respectively. 
This gives a ratio $E_k^v$(H$_2$) / $E_k^v$(D$_2$) = 1.33, somewhat 
smaller than the limit 1.41 expected for harmonic vibrations at low 
temperatures.  This deviation may be due to both anharmonicity in
the interatomic interaction and changes in the effective mass caused by
coupling to the host atoms.
This ratio decreases as $T$ is raised, and amounts to 1.19 at 900 K.
For comparison we also present in Fig.~2 the kinetic energy corresponding
to the classical limit with four vibrational degrees of freedom 
($2 k_B T$, dotted line).

\subsection{Atomic delocalization}

To study the spatial delocalization of a quantum particle from PIMD
simulations, it is convenient to consider the center-of-gravity (centroid) 
of the quantum paths of the particle, defined as
\begin{equation}
   \overline{\bf r} = \frac{1}{P} \sum_{i=1}^P {\bf r}_i  \, ,
\label{centr}
\end{equation}
${\bf r}_i$ being the coordinates of the ``beads'' in the associated ring
polymer.

The mean-square displacement of a quantum particle along a PIMD
simulation run is then given by:
\begin{equation}
\Delta^2 =  \frac{1}{P} \left< \sum_{i=1}^P 
           ({\bf r}_i - \left< \overline{\bf r} \right>)^2
           \right>    \, ,
\label{delta2}
\end{equation}
where $\langle ... \rangle$ indicates a thermal average at temperature $T$.
After some straightforward manipulations, one can write $\Delta^2$ as
\begin{equation}
 \Delta^2 = \Delta^2_Q + \Delta^2_C  \, ,
\label{delta2b}
\end{equation}
with
\begin{equation}
 \Delta^2_Q = \frac{1}{P} \left< \sum_{i=1}^P 
             ({\bf r}_i - \overline{\bf r})^2 \right>    \, ,
\end{equation}
and
\begin{equation}
  \Delta^2_C = \left< (\overline{\bf r} - 
                \left< \overline{\bf r} \right>)^2 \right>    \, .
\end{equation}
The first term, $\Delta^2_Q$, is the mean-square ``radius-of-gyration''
of the ring polymers associated to the quantum particle (atomic nucleus) 
under consideration.\cite{gi88} This is a measure of the average 
extension of the paths and, therefore, of the importance of
quantum effects in a given problem.
The second term in Eq.~(\ref{delta2b}) is the mean-square
displacement of the center of gravity of the paths.
This term is the only one surviving at high temperatures,
since in the classical limit each path collapses onto a single point
(hence with a vanishing radius-of-gyration).
For situations in which the anharmonicity is not extremely large, the
distribution of $\overline{\bf r}$ is similar to that of a classical 
particle in the same potential, and thus $\Delta^2_C$ can be considered
as a kind of semiclassical delocalization. 

\begin{figure}
\vspace{-2.0cm}
\hspace{-0.5cm}
\includegraphics[width= 9cm]{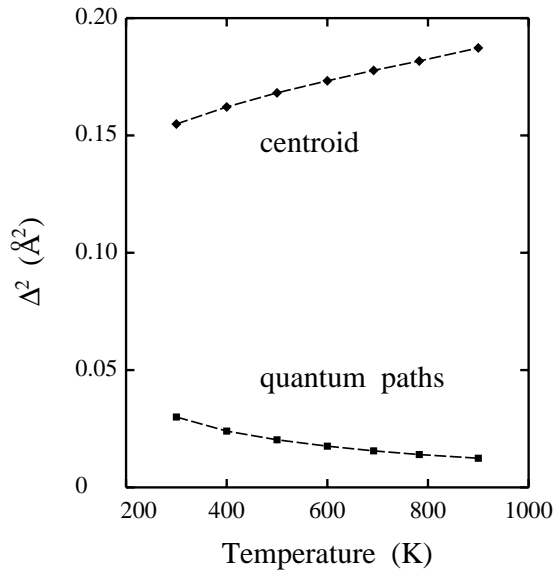}
\vspace{-2.5cm}
\caption{
Spatial delocalization of atomic nuclei (protons) in H$_2$.
Diamonds indicate the mean-square displacement of the centroid of the
quantum paths, $\Delta_C^2$, and squares correspond to the mean-square
radius-of-gyration of the paths, $\Delta_Q^2$.
}
\label{f3}
\end{figure}

Going back to our problem of H$_2$ in silicon, 
for each hydrogen atom in the molecule we have calculated separately
both terms giving the atomic delocalization in Eq.~(\ref{delta2b}).
Shown in Fig.~3 are the values of $\Delta^2_Q$ (spreading of the quantum
paths, squares) and $\Delta^2_C$ (centroid delocalization, diamonds),
as derived from our PIMD simulations at several temperatures. 
In this plot, one observes that $\Delta^2_C$ is much larger than 
$\Delta^2_Q$ in the whole temperature range under consideration.
This is not strange if one takes into account that the molecular rotation
around the interstitial T site can be considered as a classical motion
at these temperatures. In fact, the order of magnitude of this 
spatial delocalization can be obtained from the free motion of a particle
on a spherical surface with radius equal to half distance between 
atoms in an H$_2$ molecule. For a distance $R = 0.77$ \AA, we obtain
a mean-square classical displacement of 0.15 \AA$^2$, close to the 
value of $\Delta^2_C$ at 300 K. This magnitude increases for rising
temperature, as expected for an increase in the fluctuations of the
distance from each H atom to the average position (T site). 

For the spreading of the quantum paths of each H atom we obtain at
room temperature $\Delta^2_Q$ = 0.03 \AA$^2$, and it decreases as
temperature is raised.
This gives for the paths an average extension of $\sim$ 0.1 \AA\ at
300 K, much smaller than the H-H distance, and thus justifying the 
neglect of quantum exchange between protons. 
Moreover, the fact that $\Delta^2_Q$ is much smaller than $\Delta^2_C$
in the temperature range considered here does not mean that quantum
effects are irrelevant, but is a consequence of the enhancement
of $\Delta^2_C$ due to molecular rotation.

\subsection{Interatomic distance}

As mentioned above, the interatomic distance between hydrogen
atoms increases when the molecule is introduced from the gas phase into 
a silicon crystal, due to the interaction between H and host atoms.
For the minimum-energy distance we found $R_0$ = 0.752 \AA,
which is smaller that the values obtained in earlier calculations
(0.788 \AA\ in Ref.~\onlinecite{ho98} and 0.817 \AA\ in
Refs.~\onlinecite{wa98,wa99}).

\begin{figure}
\vspace{-2.0cm}
\hspace{-0.5cm}
\includegraphics[width= 9cm]{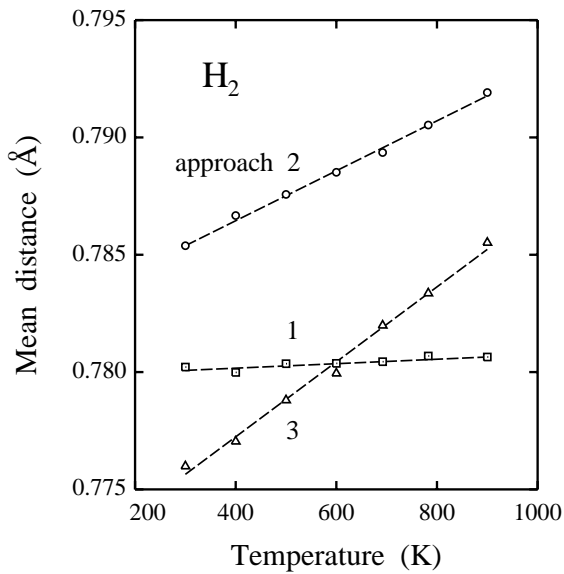}
\vspace{-2.5cm}
\caption{
Mean distance between H atoms in an H$_2$ molecule in silicon.
Symbols correspond to different approximations for the molecular motion.
Squares: motion of H$_2$ in one dimension with fixed host atoms (approach 1);
circles: free motion of H$_2$ in a fixed silicon lattice (approach 2);
triangles: free motion of H$_2$ and host atoms (approach 3).
Error bars are in the order of the symbol size.
Dashed lines are linear fits to the data points.
}
\label{f4}
\end{figure}

We now present the temperature dependence of the mean distance H--H
for the three approaches considered above to study molecular
hydrogen in silicon.
Our results are displayed in Fig.~4, where symbols represent data
points derived from PIMD simulations.
For approach 1 (1d motion in a fixed lattice), we find at 300 K
a mean distance $R$ = 0.780 \AA, which represents an appreciable
increase vs the distance obtained for the minimum-energy configuration.
In this model, $R$ increases very slowly as
a function of $T$ (squares in Fig.~4), since molecular rotation is
not allowed and the molecule expansion is only due to the increasing 
population of excited vibrational levels. In fact, we find 
$d R/d T = 1.0 \times 10^{-6}$~\AA/K.

When molecular rotation is allowed in a fixed lattice (approach 2),
we observe an increase in $R$ (see circles in Fig.~4).
At 300 K we found $R$ = 0.785 \AA, about $5 \times 10^{-3}$ \AA\
larger than for 1d motion. Now $R$ rises with temperature much faster
than in approach 1, with $d R/d T = 1.1 \times 10^{-5}$ \AA/K. 
Next, we allow the Si atoms to move, introducing the full
quantization of all degrees of freedom in the simulation cell, and
we obtain a reduction of the distance H--H with respect to approach
2. This can be understood as due to a softening of the effective Si--H 
interaction, which decreases as a consequence of Si motion. 
In this case with full motion of the 66 atoms in the cell, we find
$d R/d T = 1.6 \times 10^{-5}$ \AA/K, which means a larger slope
than in approach 2.

It is interesting to compare these changes in the mean distance $R$
with those corresponding to molecular hydrogen in the gas phase. 
To this end
we have carried out some PIMD simulations of an isolated hydrogen
molecule with the same interatomic potential. In this case we obtain an
increase in $R$ with temperature given by 
$d R/d T = 7.5 \times 10^{-6}$ \AA/$K$, a value clearly smaller than
those obtained for H$_2$ in silicon in our approaches 2 and 3.
This means that, for H$_2$ in silicon, the change of interatomic
distance with temperature is controlled by both the centrifugal
expansion due to rotation, and interaction with the nearby host atoms.

\begin{figure}
\vspace{-2.0cm}
\hspace{-0.5cm}
\includegraphics[width= 9cm]{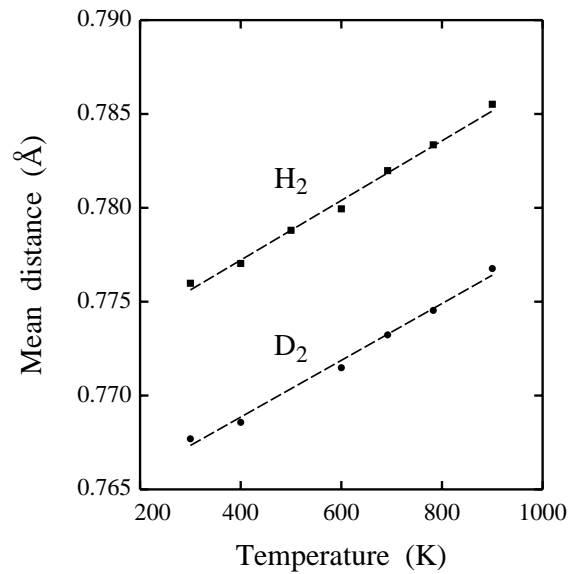}
\vspace{-2.5cm}
\caption{
Mean interatomic distance for H$_2$ and D$_2$ molecules in silicon,
as a function of temperature.
Symbols indicate results derived from PIMD simulations for approach 3,
in which all atoms are mobile:
squares for H$_2$ and circles for D$_2$.
Error bars are on the order of the symbol size.
Dashed lines are linear fits to the data points.
}
\label{f5}
\end{figure}

PIMD  simulations can be also employed to study the isotopic dependence
of the mean interatomic distance $R$. The molecular expansion with
respect to the lowest-energy classical geometry is due to a combination
of anharmonicity with quantum delocalization. One expects smaller distances 
for molecular deuterium due to its smaller vibrational amplitudes. In fact,
at 300 K we found for D$_2$ in silicon, $R$ = 0.767 \AA, to be compared
with $R$ = 0.776 \AA\ for H$_2$ at the same temperature, and a distance
$R_0$ = 0.752 \AA\ for the lowest-energy position in the classical limit.
In Fig.~5 we present the temperature dependence of the mean distance for
both H$_2$ and D$_2$, as derived from our PIMD simulations for approach 3
(full motion of molecular hydrogen and host atoms). For D$_2$ we
find $d R/d T = 1.5(1) \times 10^{-5}$ \AA/$K$, which coincides within
error bar with the slope obtained for H$_2$ in silicon in the temperature
region from 300 to 900 K.

\subsection{Stretching frequency} 

The stretching frequency of H$_2$ is an important fingerprint of 
the molecule, that in fact has been used to detect and characterize
this impurity in the silicon bulk.\cite{le98,zh99} 
This stretching vibration has been found at 3618 cm$^{-1}$ (at 4 K)
independently by Raman\cite{le98} and infrared absorption 
spectroscopies.\cite{pr98}

\begin{figure}
\vspace{-2.0cm}
\hspace{-0.5cm}
\includegraphics[width= 9cm]{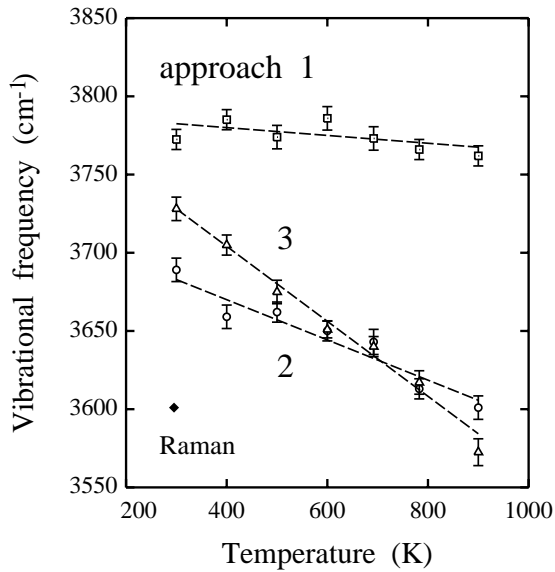}
\vspace{-2.5cm}
\caption{
Frequency of the stretching vibration of the H$_2$ molecule in silicon
as a function of temperature. Symbols represent results derived from
PIMD simulations in three different approximations:
squares, motion of H$_2$ in one dimension in a fixed silicon lattice;
circles, free motion of H$_2$ with fixed Si atoms;
triangles, free motion of H$_2$ and Si atoms.
Error bars correspond to the statistical uncertainty in the PIMD
simulations.
A black diamond indicates the stretching frequency obtained by
Raman spectroscopy at room temperature.\cite{le98}
}
\label{f6}
\end{figure}

In Fig.~6 we show the temperature dependence of $\omega_s$ for
H$_2$ in the three approaches considered here, as derived from the LR
method presented above.
In approach 1 (1d motion) the frequency decreases slightly in the
analyzed temperature range.
In approaches 2 and 3, the coupling between molecular rotation
and vibration causes an appreciable change of $\omega_s$ with the
temperature. For model 2 (fixed Si lattice) we find 
$d \omega_s / d T = - 0.13$  cm$^{-1}$/K, to be compared with a
slope of $d \omega_s / d T = - 0.24$  cm$^{-1}$/K for model 3,
which includes motion of the host atoms.  
Thus, motion of the Si atoms causes a significant change in
$d \omega_s / d T$, which becomes almost twice larger than in the
case of a static Si lattice.
It is interesting that at room temperature $\omega_s$ is smaller for
model 2 than for approach 3, but due to its faster decrease in
the latter approach, $\omega_s$ becomes smaller for model 3 at high
$T$.

Something similar has been obtained for the stretching vibration of
D$_2$. In particular, for approach 3 we find a rather constant ratio
between the stretching frequencies of H$_2$ and D$_2$, that amounts to
1.37(1), somewhat smaller than the ratio expected in a harmonic
approximation (1.41).
Experimentally, a ratio of 1.37 is found from infrared\cite{pr97} and
Raman\cite{le98,la02} spectra of H$_2$ and D$_2$ in silicon, a little 
smaller than the ratio 1.39 observed for these molecules in the gas 
phase.\cite{st57}

For the HD molecule in silicon, an infrared study allowed to
determine the energy of the first excited rotational level.\cite{ch02a}
In fact, a value of 73.9 cm$^{-1}$ was found for the wave-number
difference between the levels $J$ = 0 and $J$ = 1, somewhat lower than
that corresponding to the gas phase (89.3 cm$^{-1}$).
By scaling that wave-number difference with the reduced mass, we
expect for H$_2$ an energy difference of about 99 cm$^{-1}$.
Since our PIMD simulations yield results for the average frequency
$\omega_s$, one can estimate a frequency shift from
the rotational energy, taking into account the population and
degeneracy of the different levels.\cite{ch02a}
By considering only the levels $J$ = 0 and $J$ = 1, one would expect at 
room temperature a frequency shift $d \omega_s / d T$ on the order of 
$-0.05$ cm$^{-1}$/K, clearly lower than the value found from our simulations 
for approach 3 ($-0.24$ cm$^{-1}$/K).
This is not strange, taking into account that at these temperatures 
higher rotational levels will be excited, contributing to a larger decrease
in the average frequency.
However, the actual position of these levels further than $J$ = 1
is not known at present, and a more detailed comparison with our
results is not possible.

We note that the quantum treatment of atomic nuclei in molecular
dynamics simulations is crucial to give a reliable description of
the vibrational frequencies of light atoms like hydrogen.
In fact, we have applied the LR method to calculate the stretching
frequency $\omega_s$ from classical simulations. At 300 K we
found for H$_2$ in silicon a frequency $\omega_s$ = 4039 cm$^{-1}$ 
(for full motion of interstitial hydrogen and host atoms), to be 
compared with $\omega_s$ = 3728 cm$^{-1}$ derived from
PIMD simulations. As expected, the classical value is much closer to
the frequency $\omega_s$ = 4071 cm$^{-1}$ obtained in a HA for
H$_2$ in silicon.

\section{Discussion}

In Sec. III we have presented results of our PIMD simulations for 
H$_2$ and D$_2$ in silicon.
The main advantage of this kind of simulations is the possibility
of calculating energies at finite temperatures, with the
inclusion of quantization of host-atom motions, which are not 
easy to be accounted for in fixed-lattice calculations. 
Isotope effects can be readily explored, since the impurity mass
appears as a parameter in the calculations. This includes the
consideration of zero-point motion, which together with anharmonicity
gives rise to non-trivial effects.
In addition, the vibrational motion of H$_2$ is coupled with molecular 
rotation, leading to a change in the stretching frequency with
temperature.

As mentioned above,
an important feature of isolated H$_2$ molecules in semiconductors
is their stretching vibration $\omega_s$. In a harmonic approximation, 
the tight-binding potential employed here yields for H$_2$ in silicon 
a frequency $\omega_s$ = 4071 cm$^{-1}$ vs 4397 cm$^{-1}$ for H$_2$ in
the gas phase, which means a reduction of about 330 cm$^{-1}$ due to
interaction with the host atoms.  This reduction is accompanied by an
increase in the H--H distance, as shown in Sect.III.C. 
An additional decrease in $\omega_s$
is obtained when anharmonic effects are taken into account in a
one-dimensional motion of the molecule in a fixed lattice. In fact,
at 300 K the LR calculations give in this case 
$\omega_s$ = 3770 cm$^{-1}$, which means a decrease in frequency 
of about 300 cm$^{-1}$ with respect to the HA for H$_2$ in silicon.
This frequency change due to anharmonicity is in the order of that
derived in Ref.~\onlinecite{wa99} from DF calculations, namely
$\Delta \omega_s = -408$ cm$^{-1}$.
This frequency is further lowered when full (quantum) motion of
the molecule and host atoms are allowed, giving 
$\omega_s$ = 3728 cm$^{-1}$. In this latter reduction there is
a contribution of two competing effects: coupling 
between molecular rotation and vibration, and interaction with
Si atoms, whose motion allows for a larger delocalization of the
H$_2$ molecule in the interstitial space.

In general we observe a correlation between $\omega_s$ and mean
interatomic distance $R$ in the H$_2$ molecule, in the sense that a
rise in $\omega_s$ is accompanied by a decrease in $R$.
This is in line with the general trend found by Van de Walle\cite{wa98}
for molecular hydrogen in crystalline semiconductors, as derived from
DF calculations at $T=0$.
However, this trend is not so strict when atomic motion is included at
finite temperatures, as derived from Figs.~4 and 6. 
In this case, the mobile Si atoms may contribute to an additional decrease 
in the stretching frequency of H$_2$ by a rise in the effective mass
associated to this vibrational mode.

In connection with this, it is clear that
theoretical techniques to deal with the electronic structure of solids
have been improving their precision over the years. For various purposes, 
the accuracy currently achieved by these methods is excellent, when comparing
their predictions with experimental data. However, quantum nuclear effects 
limit the accuracy of state-of-the-art techniques to predict actual
properties of light impurities in solids. The answer to this question
has to be found in {\em ab-initio} path-integral simulations, where both
electrons and nuclei are treated directly from first principles. 
But even in this case the question is not simple when one has to deal with
phenomena such as molecular rotation at low temperatures, where a proper
description of quantum rotation has to be included in the formalism.

There is an important challenging point that should be considered in
future work. It refers to considering coupling between nuclear spins 
in the hydrogen molecule, i.e. dealing separately with ortho and 
para-H$_2$ (both have been observed in silicon\cite{ch02b,la02,hi07}). 
This becomes specially relevant at low temperatures, where
the quantum character of molecular rotation has to be explicitly
considered in the simulations. Usually this kind of calculations have
been carried out by assuming the molecule to be a rigid rotor, without
taking into account vibrations and deformations, and thus neglecting
the ro-vibrational coupling.

  An analysis of hydrogen diffusion in silicon is out of the scope 
of this paper.  Actual diffusion coefficients are not 
directly accessible with the kind of simulations employed here, since 
the time scale employed in the calculations is not readily connected 
to the real one. In this respect, PIMD simulations could be applied 
to study quantum diffusion of H$_2$ in silicon, by calculating 
free-energy barriers in a way similar to that employed earlier to study 
the diffusion of atomic hydrogen in metals\cite{ma95} and 
semiconductors.\cite{he97,he07}

In summary, the PIMD method has turned out to be well-suited to study
finite-temperature equilibrium properties of hydrogen molecules
in silicon.  This has allowed us to notice the importance of 
anharmonicity and ro-vibrational coupling in order to give a 
realistic description of the properties of these interstitial impurities. 
Anharmonicity shows up in the stretching motion of the molecules,
causing important shifts with respect to the harmonic expectancy.

\begin{acknowledgments}
This work was supported by CICYT (Spain) through Grant
No. BFM2006-12117-C04-03.
\end{acknowledgments}

\end{document}